\documentclass{aa}  

\usepackage{graphicx}
\usepackage{txfonts}
\usepackage[colorlinks=true, citecolor=blue, linkcolor=blue]{hyperref}
\usepackage{xcolor}
\usepackage[normalem]{ulem}
\usepackage{natbib}
\bibpunct{(}{)}{;}{a}{}{,} % to follow the A&A style

\begin{document}

   \title{Relation between magnetic field inclination and apparent motion of penumbral grains}

   \subtitle{}

   \author{Michal Sobotka
          \inst{1}\fnmsep\thanks{michal.sobotka@asu.cas.cz}
          \and
          Jan Jur\v{c}\'ak \inst{1}
          \and
          J.~Sebasti\'an {Castellanos~Dur\'an} \inst{2}
          \and
          Marta Garc\'{\i}a-Rivas \inst{1,3}
          }

   \institute{Astronomical Institute of the Czech Academy of Sciences,
              Fri\v{c}ova 298, 25165 Ond\v{r}ejov, Czech Republic
   \and
    Max-Planck Institute for Solar System Research, Justus-von-Liebig-Weg 3,
    D-37077, G\"ottingen, Germany
   \and
    Astronomical Institute of Charles University, Faculty of Mathematics and Physics, V Hole\v{s}ovi\v{c}k\'{a}ch 2, 180 00 Praha 8, \\ Czech Republic
 }

  \date{Received 15 September 2023; accepted 17 November 2023}

\abstract
  % context heading (optional) {} leave it empty if necessary
   {Bright heads of penumbral filaments, penumbral grains (PGs), show apparent horizontal motions inwards, towards the umbra, or outwards, away from the umbra.}
   {We aim to prove statistically whether the direction of PGs' apparent motion is related to the inclination of the surrounding magnetic field.}
   {We use spectropolarimetric observations of five sunspot penumbrae to compare magnetic inclinations inside PGs with those in their surroundings. The data are taken by three observatories: Hinode satellite, Swedish Solar Telescope, and GREGOR solar telescope. The direction of PGs' motion is determined by feature tracking. The atmospheric conditions in PGs and their surroundings, including the magnetic field information, are retrieved by means of height-stratified spectropolarimetric inversions.}
   {On a sample of 444 inward- and 269 outward-moving PGs we show that 43\% of the inward-moving PGs have magnetic inclination larger by $8^\circ \pm 4^\circ$ than the inclination in their surroundings and 51\% of the outward-moving PGs have the inclination smaller by $13^\circ \pm 7^\circ$ than the surrounding one. The opposite relation of inclinations is observed at only one-fifth of the inward- and outward-moving PGs.}
   {Rising hot plasma in PGs surrounded by a less inclined magnetic field may adapt its trajectory to be more vertical, causing an inward apparent motion of PGs. Oppositely, it may be dragged by a more horizontal surrounding magnetic field such that an outward apparent motion is observed.}


  \titlerunning{Apparent motion of penumbral grains}

\maketitle
%
%%%%%%%%%%%%%%%%%%%%%%%%%%%%%%%%%%%%%%%%%%%%%%%%%%%%%%%%%%%%%%%%%%%%%%%%

\section{Introduction}
\label{sec:intro}

A sunspot penumbra has a well-marked radial filamentary structure observed in white-light images. This structure consists of bright and dark fibrils. The bright fibrils usually have bright heads that point towards the umbra and appear all over the penumbra. \citet{Muller1973} called these bright heads penumbral grains (PGs) and reported that they move towards the umbra (inwards). In addition to inward-moving PGs, \citet{SBSIII1999} also found PGs that move away from the umbra (outwards) and, in contrast to the inward-moving PGs, are concentrated in the outer penumbra. These observations were confirmed by \citet{SSIV2001} and \citet{Zhang2013}. From the aforementioned works, it can be derived that the observed width of PGs varies typically between 0\farcs 3 and 0\farcs 5, their length between 0\farcs 6 and 2\farcs 0, and their lifetime between 10 and 50 minutes. The inward-moving PGs live slightly longer than the outward-moving ones and short-lived PGs are more numerous than the long-lived ones, where $\sim$3/4 of PGs move inwards \citep{SBSIII1999}. Spectroscopic observations show upflows of hot gas coinciding with PGs \citep{Ichimoto2007, Franz2009}.

The first theoretical explanation of inward motion of PGs (and other observed properties of penumbral filaments) was proposed by \citet{Schlich1998ApJ, Schlich1998AA}. An inclined magnetic flux tube with an upflow of hot sub-photospheric gas rises and crosses the visible surface. As the flux tube rises, its inclination to the normal decreases. At the intersection between the surface and the flux tube, a PG is observed and it shows an apparent motion towards the umbra. The part of the tube above the visible surface is horizontal and the outward-moving hot gas inside is cooled by radiation. As long as the flux tube is hotter than the surroundings, it is observed as a bright fibril. But when the gas cools down, it becomes dark. The outward-moving gas constitutes the Evershed flow \citep{Evershed1909, Rimmele2006}.

This concept has been extended in terms of magnetoconvection. Magnetohydrodynamical (MHD) simulations of sunspot penumbrae have shown that the filamentary structure is created by convective cells in highly inclined magnetic field. First simulations of the penumbra in a slab geometry \citep[e.g.,][]{Heinemann2007, Scharmer2008} produced bright filamentary structures with horizontal flows, similar to observed Evershed flows. These structures moved towards the umbra, resembling inward motions of PGs. More advanced simulations of penumbral fine structure were presented by \citet{Rempel2012}. An animation of the temporal evolution of the sunspot fine structure \citep[Fig.~6 in the online version of][]{Rempel2012} shows the inward motion of PGs in the whole penumbra, but there are also some bright structures in the outer penumbra that move outwards. We note that the apparent motions of PGs are not discussed by \citet{Rempel2012}.

Semi-empirical models of typical penumbral filaments were proposed by \citet{Tiwari2013} using Hinode spectropolarimetric observations. In these models, the hot gas from subphotospheric layers emerges in the PG, which is the hottest and brightest part of the filament. Then the gas flows along a filament body with a nearly horizontal magnetic field, constituting the Evershed flow, leaks laterally, producing local downflows on the sides of the filament \citep[cf.][]{Scharmer2011Sci, Joshi2011ApJ}, and cools down by radiation.
The brightness of the filament's body is at maximum in the part facing the umbra and decreases radially outwards. Finally, the cool gas sinks at the end of the filament, producing a strong downflow. The filaments are embedded in a less inclined surrounding magnetic field, spines \citep{Lites1993}, which is presumably a continuation of the umbral magnetic field \citep{Tiwari2015}. Its inclination to the normal gradually increases with the distance from the umbra. According to their models \citep[Fig. 6 of][]{Tiwari2013}, deep in the photosphere at optical depth $\tau_{\rm 500\,nm} = 1$, the magnetic field strength in PGs is substantially weaker than the surrounding one in the inner penumbra and comparable in the outer penumbra. The magnetic field inclination in PGs is larger (more horizontal) than the surrounding one in the inner penumbra and smaller (more vertical) in the outer penumbra. This semi-empirical model is in agreement with the MHD simulations of penumbral fine structure \citep{Rempel2012}.

\begin{table*}\centering
\caption{Summary of observations.}  \label{tab:obs}
\begin{tabular}{lccccc}
\hline\hline
Data set     & 1 & 2 & 3 & 4 & 5 \\
\hline
Date         & 2006-12-10 & 2007-05-01 & 2012-05-05 & 2017-09-03 & 2022-05-18 \\
Time UT      &11:56--12:57&21:44--22:24&08:11--09:00&08:43--09:05&08:29--08:48\\
Active region & 10930      & 10953      & 11471      & 12674      & 13014     \\
Position $\mu$ &  0.92     &  0.99      &  0.92      &  0.96      &  0.82    \\
\hline
Instrument   &  SOT--SP     &  SOT--SP    &  CRISP     &  GRIS      &  GRIS      \\
Spectral lines &\ion{Fe}{i} 630.2&\ion{Fe}{i} 630.2&\ion{Fe}{i} 617.3&\ion{Fe}{i} 1564.9&\ion{Si}{i} 1082.7 \\
~~~[nm]      & \ion{Fe}{i} 630.3 & \ion{Fe}{i} 630.3 &    -       & \ion{Fe}{i} 1566.2 & \ion{Ca}{i} 1083.9 \\
Sampling    & 0\farcs 16 & 0\farcs 16 & 0\farcs 059 & 0\farcs 135 & 0\farcs 135 \\
Inversion code &  SPINOR--2D  &  SPINOR--2D    &  SIR       &  SIR       &  SIR        \\
$\tau_{\rm 500\,nm}$ &  0.16      &  0.16      &  1.0       &  1.0       &  1.0        \\
\hline
Broadband    &  $G$ band  & continuum  & continuum  &  none      &  TiO band   \\  
~~wavelength & 430.9 nm   & 450.4 nm   &  617.4 nm  &     -      &  750.7 nm   \\
~~cadence    &  120 s     &  30 s      &  56.5 s    &     -      &  5.47 s     \\
~~sampling   & 0\farcs 109 & 0\farcs 054 & 0\farcs 059 &  -      & 0\farcs 050 \\
\hline
\end{tabular}
\tablefoot{Position $\mu$ -- cosine of the heliocentric angle; sampling -- angular size of a pixel; $\tau_{\rm 500\,nm}$ -- continuum optical depth, at which the magnetic-field parameters are retrieved from the model atmosphere.}
\end{table*}

Applying the time-slice method on a six-hour long series of high spatial (0\farcs 14) and temporal (20\,s) resolution images of a large sunspot, \citet{SobPusch2022} measured horizontal motions in the penumbra. This data set was acquired on 18~June 2004 at the 1-m Swedish Solar telescope \citep[SST,][]{SchaSST2003} under excellent seeing conditions. They found that the horizontal speed of PGs motions changed from $-0.7$\,km\,s$^{-1}$ inwards in the inner penumbra to 0.4\,km\,s$^{-1}$ outwards in the outer penumbra. The orientation of motions changed from inwards to outwards in the middle penumbra and the outward speed gradually increased with distance in the outer penumbra. \citet{SobPusch2022} suggested that the apparent motions of PGs may be affected by the inclination of the surrounding magnetic field: ``Rising hot plasma surrounded by a stronger and less inclined magnetic field may adapt its trajectory to be more vertical, which would lead to the inward motion of PGs in the inner penumbra. Oppositely, in the outer penumbra the rising hot plasma in the filament's head is dragged by the surrounding, more horizontal magnetic field such that its crossing point with the visible surface (PG) moves outwards.''

The aim of this work is to statistically test whether the magnetic-field inclination in the inward-moving PGs is really larger than that in their surroundings and, on the contrary, smaller in the outward-moving PGs. High-resolution spectropolarimetric observations of sunspot penumbrae together with time series of broadband or continuum images are used for this purpose.

%%%%%%%%%%%%%%%%%%%%%%%%%%%%%%%%%%%%%%%%%%%%%%%%%%%%%%%%%%%%%%%%%%%%%%%%

\section{Observations}
\label{sec:obs}

Magnetic field inclinations in PGs are compared to those in the surroundings using detailed inclination maps obtained from inversions of spectropolarimetric observations of magnetic-sensitive lines formed in the photosphere. The observations must be sampled with a spatial resolution better than 0\farcs 16 per pixel to resolve individual PGs. The observed sunspots should have a well-developed penumbra and not be located far from the centre of the solar disc, that is, the cosine of the heliocentric angle $\mu$ must be larger than 0.8, to avoid strong projection effects. Each spectropolarimetric observation should be accompanied by a time series of broadband or continuum images to determine the direction of PGs motions. We have collected five data sets taken by three different instruments, which meet the conditions mentioned above. Spectral lines formed at slightly different heights in the lower solar atmosphere complement each other in these observations. Table~\ref{tab:obs} summarises the observations that are sorted in chronological order.

Data sets~1 and 2 were retrieved from the MODEST catalogue of depth-dependent spatially coupled inversions of sunspots \citep{Modest2022}. This catalogue collects inversions of spectropolarimetric observations in the lines \ion{Fe}{i} 630.15 and 630.25\,nm (spectral sampling 2.15\,pm) obtained with the Hinode \citep{Kosugi2007Hino} 0.5-m Solar Optical Telescope \citep[SOT,][]{Tsuneta2008SOT} Spectropolarimeter \citep[SP,][]{Ichimoto2008SP}. Series of broadband images were downloaded from the Hinode Science Data Centre Europe\footnote{\tt http://sdc.uio.no/sdc/}. 

A large roundish sunspot in active region (AR) 10930 at $\mu = 0.92$ (data set~1) was scanned by \mbox{SOT--SP} on 12 December 2006 with spatial steps of 0\farcs 16 and the scanning period from 11:56 to 12:57 UT was utilised to form the region of interest. At the same time, $G$-band images of the spot area were recorded by SOT with spatial sampling of 0\farcs 109 per pixel and a cadence of 120\,s. Another large spot in AR\,10953, $\mu = 0.99$, (data set~2) was scanned on 1 May 2007 with the same SOT--SP spatial resolution and the used scanning period was 21:44--22:24\,UT. The scanning was accompanied by the acquisition of SOT broadband images in a blue continuum with spatial sampling of 0\farcs 054 per pixel and a cadence of 30\,s. The field of view of these images covered only two-thirds of the spot area, mainly its northern part.

Data set~3 was acquired with the dual Fabry-P\'erot Crisp Imaging Spectropolarimeter \citep[CRISP,][]{SchaCRISP2006} attached to SST. 
This data set was described in detail by \citet{Hamedivafa2016}. The observation in the lines \ion{Fe}{i} 617.33\,nm, \ion{Ca}{ii} 854.21\,nm, and H$\alpha$ took 49 minutes, starting at 8:11\,UT on 5~May 2012.
The target was a decaying sunspot in AR\,11471 at $\mu = 0.92$. Each of the 53 full-Stokes ($I, Q, U, V$) spectral scans of the line \ion{Fe}{i} 617.33\,nm consisted of images at 31 wavelength steps of 2.8\,pm, distributed between $-30.8$ and $+53.2$\,pm from the line centre. We selected the best image-quality scan for further analysis. The last scanning step in $I$ at 617.39\,nm showed practically a continuum image. Time series of 53 such images with a cadence of 56.5\,s was used to determine the directions of PGs motions. The spatial sampling was 0\farcs 059 per pixel.

Data sets 4 and 5 were observed with the GREGOR Infrared Spectropolarimeter \citep[GRIS,][]{Collados2012} attached to the 1.5-m GREGOR solar telescope \citep{Schmidt2012}. Data set~4 was retrieved from the GRIS archive provided by the Leibniz-Institut f\"ur Sonnenphysik (KIS) Science Data Centre\footnote{\tt https://archive.sdc.leibniz-kis.de/}. 
The western part of a roundish leading spot in AR\,12674 at $\mu = 0.96$ was scanned on 3 September 2017. The spatial sampling was 0\farcs 135 per pixel. Two scans were made between 08:43 and 09:05\,UT, each lasting $10^{\rm m} 37^{\rm s}$. The spectral region contained \ion{Fe}{i} lines 1564.85 and 1566.20\,nm with a spectral sampling of 4.0\,pm. We used the better-quality second scan for the inversion of these lines and, because no broadband observations were available, directions of PGs motions were determined by a visual comparison of two continuum (1566.82\,nm) images, derived from the two scans. The time difference between them was $10^{\rm m} 46^{\rm s}$ but it was still possible to identify enough PGs appearing in both images.

Data set~5 was acquired at GREGOR on 18 May 2022. A complex leading sunspot in AR\,13014 at $\mu = 0.82$ was scanned with GRIS from 08:29 to 08:48\,UT with spatial sampling of 0\farcs 135 per pixel. The spectral lines used for the further analysis were \ion{Si}{i} 1082.71\,nm and \ion{Ca}{i} 1083.90\,nm, with a spectral sampling of 1.81\,pm. A series of broadband images in the TiO band at 750.7\,nm was observed simultaneously using the Improved High-resolution Fast Imager \citep[HiFI+,][]{Denker2023}. Its spatial sampling was 0\farcs 05 per pixel and the cadence after data reduction 5.47\,s.

%%%%%%%%%%%%%%%%%%%%%%%%%%%%%%%%%%%%%%%%%%%%%%%%%%%%%%%%%%%%%%%%%%%%%%%%

\section{Data processing and analysis}
\label{sec:proc}

Data sets 1--5 come from different instruments and have different parameters (cf. Table~\ref{tab:obs}), consequently, there is no unique way how to process them. Observations from Hinode (sets 1 and 2) are reduced using the nominal Hinode/SOT--SP pipeline \citep{LitesIchi2013}, observations from SST/CRISP (set~3) are reduced using the CRISPRED pipeline \citep{delaCruz2015}, and, to reduce the GREGOR/GRIS data (sets 4 and 5), the GRISRED\footnote{\tt https://gitlab.leibniz-kis.de/sdc/gris/grisred} routines are used.

Inversions of data sets 1 and 2, already done in the framework of the MODEST catalogue, were performed using the SPINOR code \citep{Frutiger2000} that relies on the STOPRO routines \citep{Solanki1987PhDT} to solve the equations of radiative transfer of polarised light. In particular, MODEST builds-up using the spatially-coupled mode of SPINOR--2D \citep{vanNoort2012, vanNoort2013A&A}. By these means, during the inversion procedure, we remove the intrinsic effects of the telescope point spread function (PSF) and retrieve height-stratified information on the solar atmosphere. The inversions were performed with three nodes located at optical depths $\log\tau_{\rm 500\,nm}=[0, -0.8, -2.0]$. At these node positions, we retrieved temperature, magnetic-field strength $B$, inclination $\gamma$ and azimuth $\phi$, as well as the line-of-sight velocity. One node for microturbulence was used. Further details and applications of MODEST can be found in \citep{Modest2022, CastellanosDuran2023ApJ}.

The Stokes inversion based on response functions code \citep[SIR,][]{RuizCobo1992} is applied to data sets 3--5. The inversion code is set to allow the line-of-sight velocity, microturbulence, magnetic field strength, and inclination to change linearly with optical depth
(2 nodes at $\log\tau_{\rm 500\,nm}=[1, -3.8]$). The linear stratification with height makes it possible to use the SIR results at $\log\tau_{\rm 500\,nm} = 0$, because the slopes of stratifications of these physical parameters are determined mainly at optical depths where the line profiles are most sensitive to physical conditions and not at the actual positions of the nodes.
The temperature is allowed to change at three nodes at $\log\tau_{\rm 500\,nm}=[1, -1.9, -3.8]$ with spline interpolation between the nodes and the magnetic field azimuth is assumed to be constant with optical depth.

After the inversion of all data sets, the $180^\circ$ azimuthal ambiguity is resolved by assuming radial orientation of the magnetic field in the penumbral filaments and the values of the magnetic-field inclination and azimuth are transformed from the line-of-sight frame to the local reference frame (LRF) using routines from the AZAM code \citep{Lites1995}.

According to the semi-empirical model of penumbral filaments proposed by \citet{Tiwari2013}, magnetic signatures of penumbral grains are most conspicuous deep in the photosphere, below the continuum optical depth $\tau_{\rm 500\,nm} = 0.1$. For this reason, the maps of magnetic field strength, inclination, and azimuth are retrieved from the SPINOR-2D modelled atmospheres at $\tau_{\rm 500\,nm} = 0.16$ and from the SIR models at $\tau_{\rm 500\,nm} = 1$. The middle-node height of the SPINOR-2D inversions was selected because the information from the \ion{Fe}{I} pair at 630 nm is better constrained at this optical depth \citep[cf.] []{CastellanosDuran2020ApJ}.
We note that SPINOR-2D accounts for the smearing by the spatial PSF. As a result, the SPINOR--2D maps at $\tau_{\rm 500\,nm} = 1$ are qualitatively equal to those at $\tau_{\rm 500\,nm} = 0.16$, but seemingly noisier \citep[see examples of this effect in][]{Modest2022}. 
The magnetic inclinations are brought to a unified scale \mbox{$0^\circ$--$180^\circ$}, that is, when the dominant polarity of a sunspot is negative, $\gamma = 180^\circ - \gamma_0$ and $\phi = \phi_0 - 180^\circ$, where $\gamma_0$ and $\phi_0$ are original values. Continuum maps derived from spectral scans are used to identify PGs.

Time series of broadband images taken simultaneously with the spectral scans are used to determine the directions of PGs' motions. The TiO-band images in data set~5 are processed by the sTOOLS software package \citep{Kuckein2017stools} with a frame selection routine that selects the best-quality image of 500 frames taken in a period of $\sim$5.5\,s \citep{Denker2018SoPh}. If necessary, rigid alignment (\mbox{data sets 1, 2, 5}) and a destretching together with subsonic filtering (data set~5) are applied to the time series to remove image drift due to inaccurate pointing and image deformation caused by the seeing.

\begin{figure*}[h!]\centering
\includegraphics[width=0.442\textwidth]{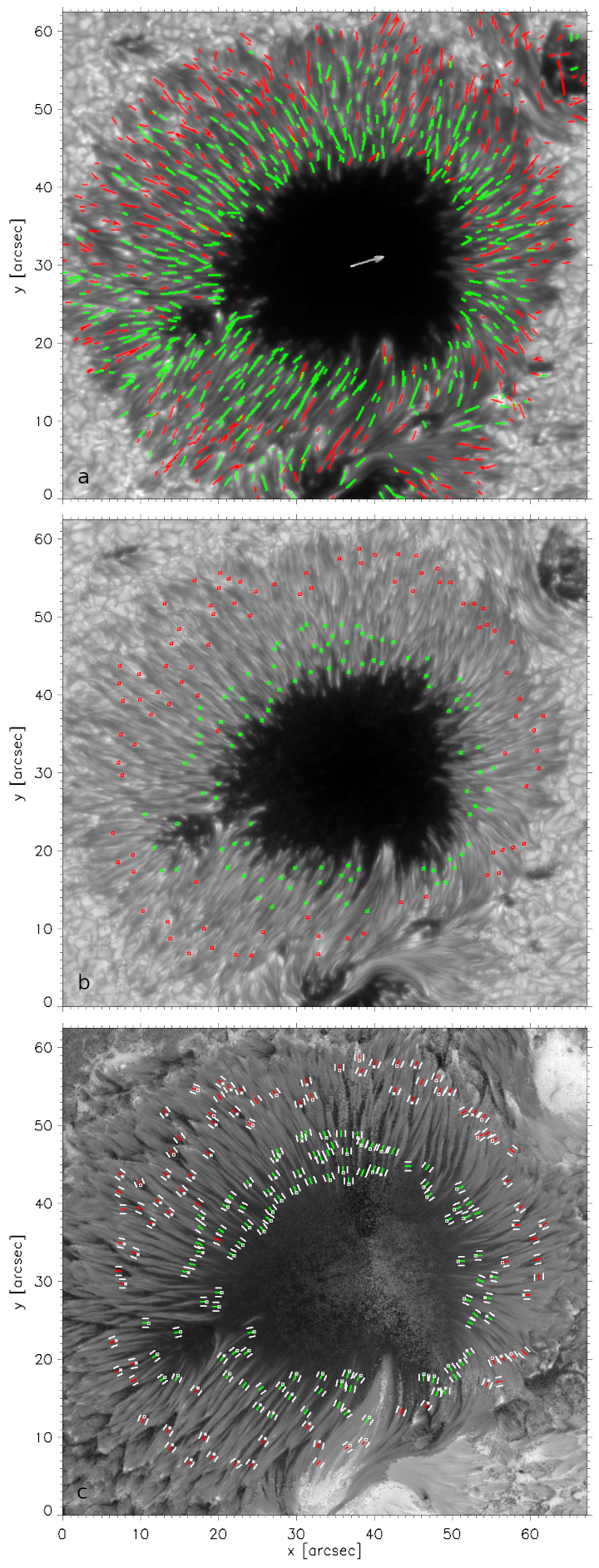}
\includegraphics[width=0.528\textwidth]{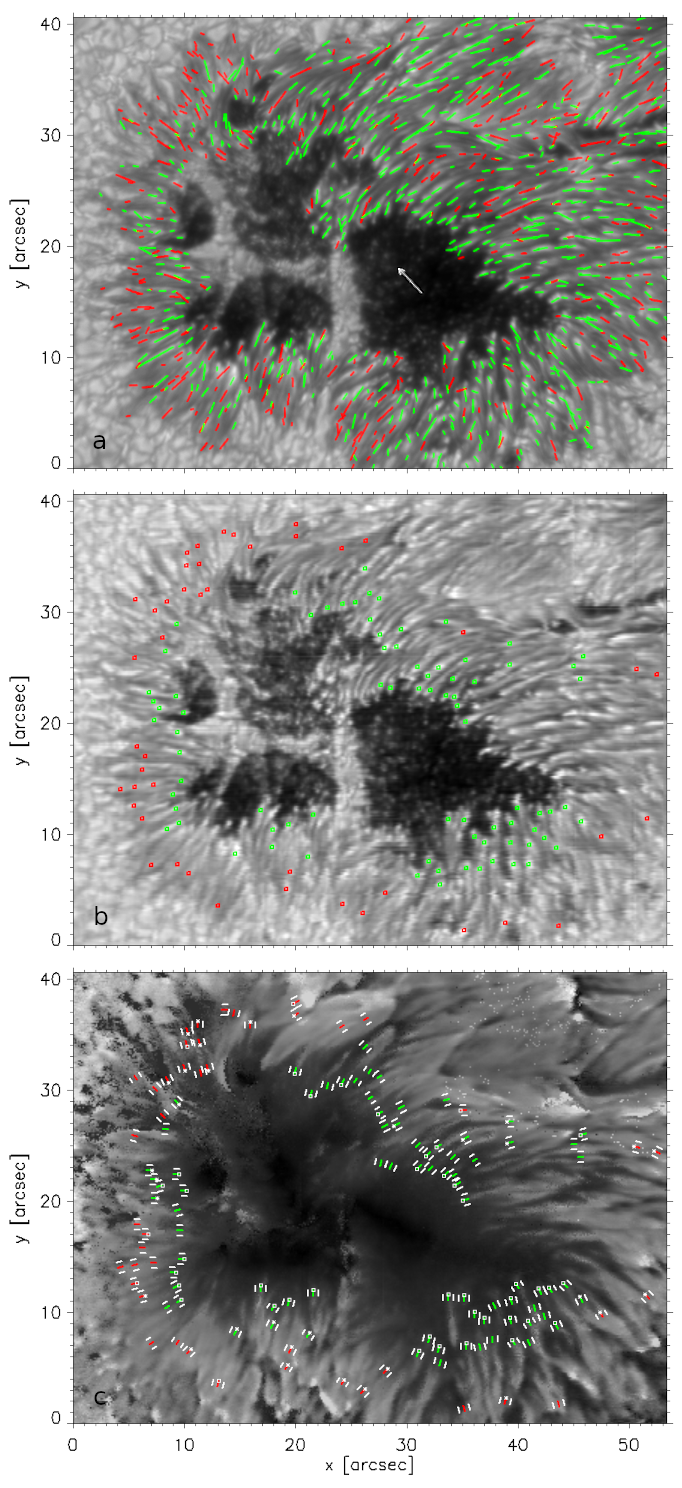}
\caption{{\it Left:} data set~1 (AR\,10930); {\it right:} data set~5 (AR\,13014). From {\it top} to {\it bottom:} (a) Trajectories of inward-moving (green) and outward-moving (red) PGs found by feature tracking from the series of broadband images. Arrows point to the disc centre. (b) Positions of visually selected PGs (green, red) in the enhanced continuum images. (c) Maps of LRF magnetic inclination ($0^\circ$--$180^\circ$) together with line segments (green, red) representing PGs. White lines show locations where the surrounding inclination was measured. Square symbols at the ends of green/red line segments mark PGs class~1, dots class $-1$.
\label{fig:Monika}}
\end{figure*}

The apparent motions of PGs are determined using the feature tracking method \citep{SBSI1997}.
Each series of broadband images is masked using a binary mask that allows for the penumbra and removes other parts of the field of view. To isolate PGs from other penumbral structures, a segmentation algorithm based on a search for regions with convex intensity profiles is applied to individual images in the series. The resulting segmentation mask is cleaned from noise using the opening operator and then it is multiplied by the original image to preserve the intensities of individual PGs. The feature-tracking procedure records the brightness, position, {\bf area}, and lifetime of PGs larger than 9~pixels (16~pixels in data set~5). The trajectories of PGs living longer than 8 minutes (2 minutes in data set~5) are reconstructed from the PGs' intensity maxima positions. The inward (INW) and outward (OUT) motions of PGs are distinguished as follows: The PGs move inwards when their trajectories start farther and end closer to the estimated geometrical centre of the sunspot and move outwards when the opposite is the case. Examples of the trajectories (green -- INW, red -- OUT) are shown in Figure~\ref{fig:Monika}a, top row, for the data sets 1 (left) and 5 (right).

%----------------------------------------------------------
%--------------------------------------------------------

At first glance, it would seem that the positions of INW and OUT PGs determined by feature tracking can be directly transferred into magnetic-inclination maps to compare $\gamma$ inside PGs and in their surroundings. However, this approach is not possible because the structures in the spectropolarimetric scans and in the series of broadband images are not synchronised in time. Different parts of the scans are taken at different times. Furthermore, several PGs may be tracked at different times in similar locations in the broadband images. We can only associate the magnetic-inclination map to those PGs that were present when the given part of the map was scanned. This problem can be solved by visual identification of PGs in continuum maps derived from the spectral scans and the use of feature-tracking results merely to distinguish the INW PGs from the OUT PGs. A local correlation tracking \citep[LCT,][]{NovSim1988} was also used for this purpose as a complementary method.

We select the PGs visually in continuum images magnified two times (except for data set~3 where the spatial sampling is more than twice as fine as in the rest of the data sets) and enhanced by unsharp masking. The selected PGs must be clearly recognised as bright heads of penumbral filaments. Moreover, they must be located in regions with a well-defined direction of PGs' motion to differentiate between the INW and OUT PGs and the local azimuth $\phi$ must be oriented parallel to the local direction of penumbral filaments. In the data set~4, where the broadband observations are missing, the PGs' motions are estimated by comparison of PGs' positions in the two continuum maps (cf. Section~\ref{sec:obs}). Positions of selected PGs (green -- INW, red -- OUT) are shown in Figure~\ref{fig:Monika}b (middle row) for the data sets 1 and 5. The inward-moving PGs are located mostly in the inner penumbra while the outward-moving ones in the outer penumbra. Total numbers of the selected INW and OUT PGs in each data set are listed in Table~\ref{tab:tabvel}.

The magnetic field inclination in PGs is retrieved from the LRF inclination maps. Here the PGs are represented by line segments with a length $l$ and centres at the positions selected in continuum images. The orientation of the line segments is determined by the local magnetic-field azimuth. The inclination $\gamma_{\rm PG}$ is the mean of values along the line segment in the inclination map. The inclination $\gamma_{\rm s}$ in the surroundings of each PG is measured along two parallel lines of the same length $l$ placed on opposite sides at a distance $d$ from the PG line segment. Inclination maps of data sets 1 and 5 together with the positions of line segments (INW PGs -- green, OUT PGs -- red, surroundings -- white) are depicted in Figure~\ref{fig:Monika}c, bottom row. We can see in the figure that the selected PGs mostly coincide with ends of elongated regions with increased inclination, which is consistent with the models of filaments proposed by \citet{Tiwari2013}.

The mean inclinations along the two surrounding lines are compared to $\gamma_{\rm PG}$. We define three classes used in this comparison. Class $-1$: $\gamma_{\rm PG} < \gamma_{\rm s}$, i.e., the inclination in a PG is smaller than the inclinations in the surroundings on both sides (the magnetic field is more vertical in the PG than in the surroundings). Class 0: unsolved cases, when $\gamma_{\rm PG}$ lies between the values of inclination on the sides or it is equal to one of them or to both. Class 1: $\gamma_{\rm PG} > \gamma_{\rm s}$, i.e., the inclination in a PG is larger than the inclinations in the surroundings on both sides (the magnetic field is more horizontal in the PG than in the surroundings).

%%%%%%%%%%%%%%%%%%%%%%%%%%%%%%%%%%%%%%%%%%%%%%%%%%%%%%%%%%%%%%%%%%%%%%%%

\section{Results}
\label{sec:results}

We study the frequencies of occurrence of the INW and OUT PGs in the three classes of magnetic inclination defined above. According to the previous works \citep[e.g.,][]{Muller1973, SBSIII1999, Zhang2013}, the length of PGs is 0\farcs 6--$2''$ and their width 0\farcs 3--0\farcs 5. These values can be used for an initial estimate of the parameters $l$ and $d$ introduced in Section~\ref{sec:proc}. Because the typical size of structures in the inclination maps varies from one data set to another, the parameters $l$ and $d$ have been refined to minimise the number of unsolved cases (class~0). Numbers of PGs falling into the three classes, together with the used values of the parameters, are listed in Table~\ref{tab:tabvel} for each data set.
In Figure~\ref{fig:Monika}c, PGs belonging to class~1 (class~$-1$) are marked by small squares (dots) at the inner ends of the green or red line segments, respectively. The absence of symbols means class~0.

\begin{table}\centering
\caption{Numbers of penumbral grains (PGs) in classes $-1$, 0, 1.}  \label{tab:tabvel}
\begin{tabular}{lccccccc}
\hline\hline
PGs  & Data & \multicolumn{2}{c}{Parameters} & \multicolumn{4}{c}{\#PGs in class} \\
type & set & $l$ & $d$ & Total & $-1$ & 0 & 1 \\
\hline
INW & 1   & 0\farcs 80 & 0\farcs 56 & 102 &  25 &  27 &  50  \\
    & 2   & 0\farcs 64 & 0\farcs 40 & 113 &  22 &  42 &  49  \\
    & 3   & 0\farcs 59 & 0\farcs 47 &  97 &  24 &  39 &  34  \\
    & 4   & 0\farcs 54 & 0\farcs 47 &  52 &  14 &  18 &  20  \\
    & 5   & 0\farcs 54 & 0\farcs 47 &  80 &  12 &  29 &  39  \\
    & All &      -     &      -     & 444 &  97 & 155 & 192  \\
\hline
OUT & 1   & 0\farcs 80 & 0\farcs 56 &  90 &  48 &  21 &  21  \\
    & 2   & 0\farcs 64 & 0\farcs 48 &  48 &  25 &  14 &   9  \\
    & 3   & 0\farcs 59 & 0\farcs 47 &  63 &  30 &  22 &  11  \\
    & 4   & 0\farcs 54 & 0\farcs 47 &  24 &  13 &   8 &   3  \\
    & 5   & 0\farcs 54 & 0\farcs 47 &  44 &  20 &  18 &   6  \\
    & All &      -     &      -     & 269 & 136 &  83 &  50  \\
\hline
\end{tabular}
\tablefoot{INW -- inward-moving PGs; OUT -- outward-moving PGs; \mbox{$l$ -- length} of a line segment representing a PG, along which the mean inclination $\gamma_{\rm PG}$ is measured; \mbox{$d$ -- distance} from a PG, at which the mean surrounding inclination $\gamma_{\rm s}$ is measured. Class~$-1$: $\gamma_{\rm PG} < \gamma_{\rm s}$; class~0: unsolved cases; class~1: $\gamma_{\rm PG} > \gamma_{\rm s}$.}
\end{table}

It can be seen from Table~\ref{tab:tabvel} that the inward-moving PGs fall most frequently into class~1 and least frequently into class~$-1$. This means that the cases where the magnetic inclination in INW PGs is larger than that in the surroundings are statistically dominant. Unresolved cases (class~0) dominate in the data set~3 but the number of class~1 PGs is still larger than that of class~$-1$ ones. The outward-moving PGs fall most frequently into class~$-1$ and least frequently into class~1, so that the statistically dominant situation is when the magnetic inclination in OUT PGs is smaller than that in the surroundings.

\begin{figure}[t]\centering
\includegraphics[width=0.49\textwidth]{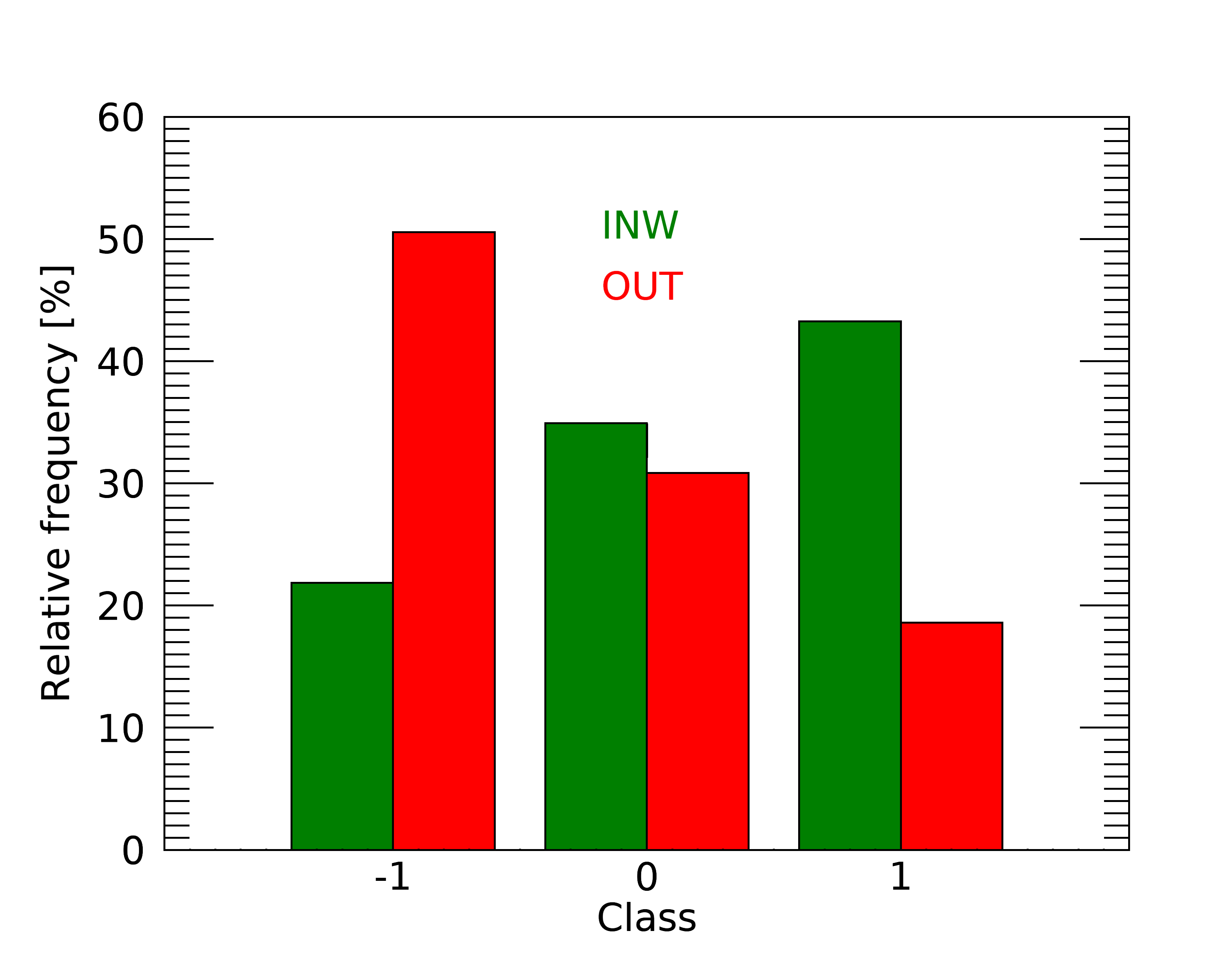}
\caption{Comparison of the magnetic-field inclinations $\gamma_{\rm PG}$ of all 713 PGs to the surrounding inclinations $\gamma_{\rm s}$. Class~$-1$: $\gamma_{\rm PG} < \gamma_{\rm s}$; class~0: unsolved cases; class~1: $\gamma_{\rm PG} > \gamma_{\rm s}$. Green -- inward-moving PGs; red -- outward-moving PGs.
\label{fig:barplot}}
\end{figure}

Histogram of relative frequencies of occurrence in the three classes for the whole sample of INW and OUT PGs is plotted in Figure~\ref{fig:barplot}. Of 444 inward-moving PGs, 43.3\% have the magnetic inclination larger than that in their surroundings, 21.8\% smaller, and 34.9\% are unsolved cases. Of 269 outward-moving PGs, 50.5\% have the inclination smaller than that in their surroundings, 18.6\% larger, and 30.9\% unknown. We can conclude that approximately a half of observed PGs comply with the hypothesis proposed by \citet{SobPusch2022}, where $\gamma_{\rm PG} > \gamma_{\rm s}$ for INW PGs and $\gamma_{\rm PG} < \gamma_{\rm s}$ for OUT PGs, while only approximately one fifth of the observed PGs are inconsistent with that.

\begin{figure}\centering
\includegraphics[width=0.49\textwidth]{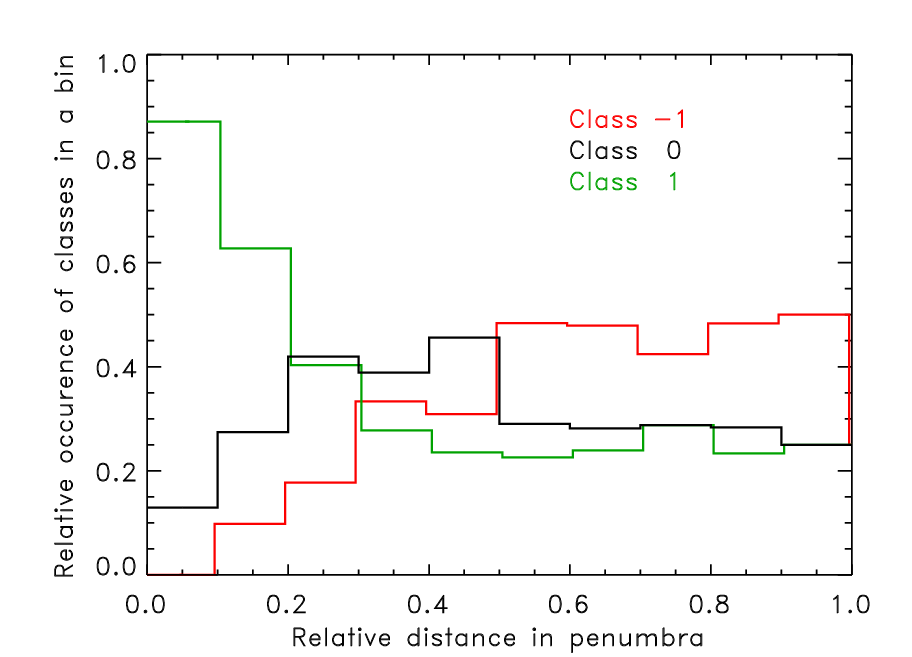}
\caption{Histogram of relative occurrences of PG classes at different relative distances in the penumbra (0: inner penumbral boundary, 1: outer penumbral boundary). Colours mark class~${-1}$ (red), class~0 (black), and class~1 (green)
\label{fig:classdcen}}
\end{figure}

Individual PG class distributions based on radial location within the penumbra disregarding the direction of motion is shown in Figure~\ref{fig:classdcen}. Data sets 1--4 obtained for approximately regular sunspots are used to make the histogram. The distances of PGs from the umbra-penumbra boundary in each individual penumbra are normalised to a common relative scale from 0 to 1 (inner to outer penumbral boundary). It can be seen that class~1 PGs are concentrated mainly in the inner penumbra, class~$-1$ in the outer penumbra, and class~0 (unsolved cases) are distributed everywhere, but most frequently in the middle penumbra. This is consistent with the semi-empirical models of penumbral filaments by \citet{Tiwari2013}.

The inward-moving PGs complying with $\gamma_{\rm PG} > \gamma_{\rm s}$ (class~1) are observed mainly in the inner penumbra where, according to the LRF inclination maps, the surrounding magnetic inclination $\gamma_{\rm s}$ varies between $35^\circ$ and $65^\circ$ and the INW PGs are, on average, more inclined by $8^\circ \pm 4^\circ$ with respect to the surrounding inclination. The outward-moving PGs complying with $\gamma_{\rm PG} < \gamma_{\rm s}$ (class~$-1$) appear mainly in the outer penumbra with $\gamma_{\rm s}$ being in the range $65^\circ$--$90^\circ$ and they are less inclined by $-13^\circ \pm 7^\circ$ on average. The minority classes, class~$-1$ for INW PGs and class~1 for OUT PGs, are observed in regions with $\gamma_{\rm s}$ between $60^\circ$ and $95^\circ$ and their average differences in inclinations $\gamma_{\rm PG} - \gamma_{\rm s}$ are $-12^\circ \pm 7^\circ$ and $10^\circ \pm 6^\circ$, respectively. In spite of diverse instruments, observing parameters, spectral lines, and inversion codes, the individual results obtained from data sets 1--5 are similar each to other and are well represented by the above values.

%%%%%%%%%%%%%%%%%%%%%%%%%%%%%%%%%%%%%%%%%%%%%%%%%%%%%%%%%%%%%%%%%%%%%%%%%%

\section{Discussion and conclusions}
\label{sec:discuss}

\begin{figure}[t]\centering
\includegraphics[width=0.49\textwidth]{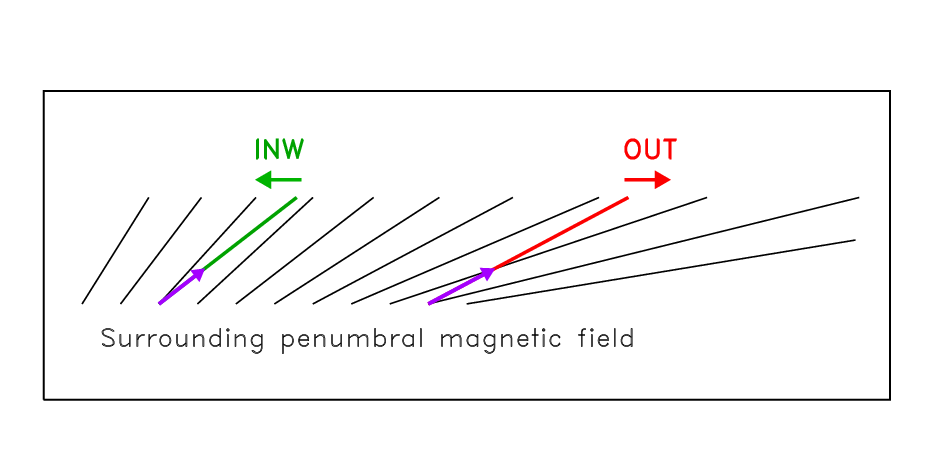}
\caption{Sketch of statistically dominant inclinations of inward-moving (green) and outward-moving (red) PGs and the inclination of the surrounding penumbral magnetic field. Green and red arrows show the directions of apparent motions of PGs and magenta arrows depict rising flows of gas.
\label{fig:sketch}}
\end{figure}

Penumbral filaments are observed everywhere in the penumbra and their bright heads, PGs, host upflows of hot gas from sub-photospheric layers. \citet{Tiwari2013} found that the filaments have essentially the same structure, particularly in the PGs, which are places where the magnetic inclination along the filament is the most vertical, around $50^\circ$. Nevertheless, this is valid only for the filaments themselves, because conditions in their surroundings vary substantially with radial distance from the umbra. In particular, $\gamma_{\rm s}$ increases from approximately $35^\circ$ at the umbra-penumbra boundary to about $90^\circ$ at the penumbra-granulation border. This increase is weaker for $\gamma_{\rm PG}$ according to our observations, as seen from the average differences $\gamma_{\rm PG} - \gamma_{\rm s} = 8^\circ$ in the inner penumbra and $-13^\circ$ in the outer one. This is illustrated in a schematic sketch in Figure~\ref{fig:sketch}. Our results are consistent with the \citeauthor{Tiwari2013}'s picture of penumbral filaments and extend it, including the apparent motion of PGs.

We utilise five data sets of high-resolution spectropolarimetric observations obtained with one space-born and two large ground-based telescopes to examine the relation between the magnetic field inclination in and around PGs and the direction of the apparent motion of PGs. \citet{SobPusch2022} suggested (cf. Section~\ref{sec:intro}) that when the magnetic inclination of a PG is more horizontal than that in its surroundings, the PG moves inwards, towards the umbra, and when it is more vertical, the PG moves outwards, away from the umbra (Figure~\ref{fig:sketch}). We find that approximately one half of 713 PGs under study really move according to this scenario.
On the other hand, approximately one fifth of PGs show opposite directions of motions and in approximately 30\% of PGs, the relation of magnetic inclinations in PGs and their surroundings is unclear.

There may be several reasons for this uncertainty. In some cases, the spatial resolution may not suffice to discern subtle variations of Stokes parameters in and around PGs, particularly in the inner penumbra where magnetic structures are packed more tightly than in the outer one. In the middle penumbra, the difference of magnetic inclinations in and around PGs is expected to be small and may be unresolved (cf. Figures~\ref{fig:classdcen} and \ref{fig:sketch}). Moreover, the accuracy of the inversion results, among them $\gamma$ and $\phi$, depends on the number of used spectral lines. For this reason, data set~3 has a large number of class 0 (unsolved) INW PGs because only one photospheric line is available unlike two lines in the other data sets. Observations at the 4-metre class telescopes (Inouye Solar Telescope and European Solar Telescope) should provide more data with better spatial resolution and polarimetric accuracy.

Our observational results show a statistical relation between the direction of apparent motions of PGs, which are locations of rising hot gas from sub-photospheric layers, and the inclination of the surrounding magnetic field in the sunspot penumbra. Numerical simulations of penumbral convection are needed to verify that.

\begin{acknowledgements}

We thank the anonymous referee for valuable comments, which helped us to improve the paper.
This work was supported by the Czech Science Foundation and Deutsche Forschungsgemeinschaft under the common grant 23-07633K and the institutional support ASU:67985815 of the Czech Academy of Sciences.
GREGOR observations and the GRIS archive were supported by SOLARNET project that has received funding from the European Union Horizon 2020 research and innovation programme under grant agreement no 824135.
The \mbox{1.5-metre} GREGOR solar telescope was built by a German consortium under the leadership of the Leibniz Institut f\"ur Sonnenphysik (KIS) in Freiburg with the Leibniz-Institut f\"ur Astrophysik Potsdam, the Institut f\"ur Astrophysik Göttingen, and the Max-Planck Institut f\"ur Sonnensystemforschung in G\"ottingen as partners, and with contributions by the Instituto de Astrof\'\i sica de Canarias and the Astronomical Institute of the  Czech Academy of Sciences.
The Swedish 1-m Solar Telescope is operated on the island of La Palma by the Institute for Solar Physics of Stockholm University in the Spanish Observatorio del Roque de los Muchachos of the Instituto de Astrof\'{\i}sica de Canarias.
Hinode is a Japanese mission developed and launched by ISAS/JAXA, with NAOJ as domestic partner and NASA and STFC (UK) as international partners. It is operated by these agencies in co-operation with ESA and NSC (Norway).
\end{acknowledgements}

%%%%%%%%%%%%%%%%%%%%%%%%%%%%%%%%%%%%%%%%%

\bibliographystyle{aa}
\bibliography{bibliography2}

\end{document}